\documentclass[runningheads]{llncs}
\usepackage[english]{babel}
\usepackage[T1]{fontenc}
\usepackage{lmodern}
\usepackage{graphicx} % Required for inserting images
\usepackage{amsmath}
\usepackage{listings}
\usepackage{xcolor}
\usepackage{rotating} % This package is necessary for rotated column names
\usepackage{array} % For defining a new column type
\newcolumntype{P}[1]{>{\centering\arraybackslash}p{#1}}
\usepackage{tabularx}
\usepackage{multirow}
\usepackage{booktabs}
\usepackage{lscape}
\usepackage{lipsum}
\usepackage{amsfonts}
\usepackage[normalem]{ulem}
\useunder{\uline}{\ul}{}
\usepackage{hyperref}
\hypersetup{
    colorlinks=false,
    linkcolor=cyan,
    filecolor=cyan,      
    urlcolor=cyan,
    pdfpagemode=FullScreen,
    }

\begin{document}

\title{TWOLAR: a TWO-step LLM-Augmented distillation method for passage Reranking}
%\thanks{Supported by organization x.}
%
\titlerunning{TWOLAR}
\author{Davide Baldelli\inst{1,3}\orcidID{0009-0006-4336-923X} \and 
Junfeng Jiang\inst{2}\orcidID{0000-0002-3680-2465} \and
Akiko Aizawa\inst{3}\orcidID{0000-0001-6544-5076} \and
Paolo Torroni\inst{1}\orcidID{0000-0002-9253-8638} }
\authorrunning{D. Baldelli et al.}
% \authorrunning{Anonymous Author(s)}
\institute{University of Bologna, Bologna, Italy \and University of Tokyo, Tokyo, Japan \and
National Institute of Informatics, Tokyo, Japan}
\maketitle              % typeset the header of the contribution
\begin{abstract}
In this paper, we present TWOLAR: a two-stage pipeline for passage reranking based on the distillation of knowledge from Large Language Models (LLM). TWOLAR introduces a new scoring strategy and a distillation process consisting in the creation of a novel and diverse training dataset. The dataset consists of 20K queries, each associated with a set of documents retrieved via four distinct retrieval methods to ensure diversity, and then reranked by exploiting the zero-shot reranking capabilities of an LLM. Our ablation studies demonstrate the contribution of each new component we introduced. Our experimental results show that TWOLAR significantly enhances the document reranking ability of the underlying model, matching and in some cases even outperforming state-of-the-art models with three orders of magnitude more parameters on the TREC-DL test sets and the zero-shot evaluation benchmark BEIR. To facilitate future work we release our data set, finetuned models, and code\footnote{Code: \url{https://github.com/Dundalia/TWOLAR}; \\
Models and Dataset: \url{https://huggingface.co/Dundalia}.}
\keywords{Information Retrieval \and Reranking \and Knowledge distillation \and Large Language Model.}
\end{abstract}
%
%
%

%Test significanza contro il miglior modello dopo il mio, aggiungere come frase, va bene anche se il risultato è negativo == NON SIGNIFICATIVO

%Agiungere a ablation study scoring strategy perche non ho messo softmax (non si allena con ranknet loss) == DONE

%Aggiungere cappello sezione 5 che prima presentiamo dati su benchmark gli ablation study sono piu tardi (per la questione che il mio omdello è inizializzato come flan t5 e non t5) == DONE

%"And obtaining results similar or even outperforming state-of-the-art models with threee orders of magnitude more parsmeters" == DONE

%Commento su ablation study su effectiveness della source ofd supervision: BM25 è ovviamente meglio perche il test è fatto cosi. Togliere interestingly e aggiugnere "this may be due to the fact" == DONE

%[BEIR CAPTION] We did test .. and obtained p-value of ..

% HUGGING FACE E REPO VERA INVECE CHE REPO ANONIMIZZATA E DRIVE

\section{Introduction}
\label{sec:Introduction}

Text ranking, a foundational task in search engines and question-answering systems, involves ordering textual documents based on their relevance to a given query or context. 
% However, while these models achieved impressive results, they also required substantial computational resources, making them less efficient compared to traditional supervised methods.

The state-of-the-art text rerankers are traditional cross-encoders like mono\-BERT~\cite{monobert}, monoT5~\cite{monot5,monot5v2}, and RankT5~\cite{RankT5}, and more recently rerankers based on Large Language Models (LLMs) like RankGPT~\cite{rankgpt} and PRP~\cite{prp}.

Cross-encoders are computationally efficient but rely on costly human-an\-no\-tat\-ed labels for training, while LLM-based methods bypass the need of in-domain fine-tuning. However, a significant downside is their substantial size and computational footprint, which could render them unsuitable for real-time inference. However, the knowledge of an LLM could be distilled to produce a student model with performance comparable to the teacher model, but a size several orders of magnitude smaller.

In this paper, we present \textbf{TWOLAR}, a \textbf{TWO}-step \textbf{L}LM-\textbf{A}ugmented distillation method for passage \textbf{R}eranking. The distillation consists in exploiting the capabilities of an LLM as a reranker to produce high-quality annotations. The annotations are applied to a dataset of queries generated artificially, either as cropped sentences or again by a specialized language model. In this way, we obtain a compact model that ranks among top performing supervised, zero-shot, and LLM-based distillation methods in various popular benchmarks.

The paper is structured as follows: Section \ref{sec:Background} provides background on text ranking methods and benchmarks. Section \ref{sec:Approach} details our approach, subdivided into scoring and distillation strategies. Section \ref{sec:Experimental setup} covers the experimental setup, including datasets, training, baselines, and results. Section \ref{sec:Discussion} discusses the results and ablation studies. Section \ref{sec:Conclusion} concludes the paper.

\section{Background}
\label{sec:Background}

Formally, given a query and a passage from a large text collection, text ranking requires returning a ranked list of the $n$ most relevant texts according to the relevance scores of a model.

The early text ranking methods relied mainly on statistical lexical features, like BM25~\cite{bm25} and TF-IDF, and used heuristic methods for retrieval~\cite{dtrsurvey}. Based on this scheme, each query-document relevance score is computed according to the lexical similarity. Subsequently, statistical language modeling has been widely explored for text ranking~\cite{statlangmodel}. With the development of machine learning, supervised approaches, which still rely on hand-crafted features as well as lexical features, have been proposed~\cite{learningtorank,learningtoranknlp}. Further progress was made with the adoption of neural networks mapping pieces of text into low-dimensional vectors to obtain better representations~\cite{adhocretr,neuralmodels4ir,deeplook}. Recently, a new paradigm emerged~\cite{pretre4ir,pretretextrank,dtrsurvey,multistage,llm4ir}, consisting of multiple stages: using a first-stage retriever that aims to reduce the candidate space by retrieving a subset of relevant candidates, often numbering in the hundreds or thousands, and then refining these initial results with a second-stage reranker.

The advent of pretrained language models (PLMs)~\cite{bert,t5,gpt2} and large-scale human annotated datasets~\cite{MSMARCO,nq,hotpotqa} marked a significant advancement in the field. Models such as DPR~\cite{dpr}, SPLADE~\cite{splade,spladev2,splade++}, and DRAGON~\cite{DRAGON} emerged as powerful first-stage retrievers. Complementing these retrievers, models like monoBERT~\cite{monobert}, monoT5~\cite{monot5,monot5v2}, and RankT5~\cite{RankT5} have been conceived as second-stage rerankers, designed explicitly to refine and optimize the results provided by the initial retrieval stage.

Recently, large language models (LLMs) have begun to play an influential role in text reranking~\cite{llm4ir,listwisellm}. The latest approaches in text reranking utilize LLMs in various ways. For instance, InPars~\cite{inpars} and InParsV2~\cite{inparsv2} leverage GPT-3 Curie~\cite{gpt3} and GPT-J~\cite{gpt-j} respectively, for data augmentation, generating synthetic queries to adapt a reranking model to different reranking tasks and domains. Other approaches instead consist in directly prompting the LLM to permute a set of documents given a query. In this vein, RankGPT~\cite{rankgpt}, LRL~\cite{listwisellm}, and PRP~\cite{prp} have demonstrated the potential of LLMs as zero-shot rerankers. Moreover, RankGPT~\cite{rankgpt} demonstrates how the ranking abilities of ChatGPT could be distilled into a more efficient DeBERTa~\cite{debertav3}. For a comprehensive survey on LLMs for Information Retrieval, the interested readers can refer to~\cite{llm4ir}.

As for the benchmarks, The largest annotated dataset for information retrieval is the MS MARCO passage reranking dataset~\cite{MSMARCO}. It contains around 530K train queries and 6.8K `dev' queries. The corpus is composed of more than 8.8M passages. For each query, relevant passages are annotated as 1 and others are annotated as 0. TREC-DL2019~\cite{trec2019} and TREC-DL2020~\cite{trec2020} are standard benchmarks derived from MS MARCO that provide dense human relevance annotations for each of their 43 and 54 queries. 
BEIR~\cite{BEIR} is a heterogeneous benchmark containing 18 retrieval datasets, covering different retrieval tasks and domains, designed for zero-shot evaluation. 

We believe that the potential of LLM distillation for text ranking has not been fully exploited yet. Our contribution aims at bridging this gap by the methodological construction of a training dataset. 

\section{Approach}
\label{sec:Approach}

Our reranking model is based on Flan-T5~\cite{flant5}, which is a text-to-text model developed as an instructed version of T5~\cite{t5}.
For our task, we use the following input template:
\begin{center}
    Query: [$Q$] Document: [$D$] Relevant:
\end{center}
where [$Q$] and [$D$] are the query and document texts, respectively, similar to the one adopted in monoT5~\cite{monot5,monot5v2}.

\subsection{Scoring Strategy} 

Flan-T5 can be straightforwardly applied to various tasks due to its text-to-text nature, such as summarization, translation, and classification. However, adapting to the ranking task is not trivial, because for each query-document pair, we usually ask models to answer with a score representing the degree of relevance. The state-of-the-art rerankers, monoT5~\cite{monot5,monot5v2} and RankT5~\cite{RankT5}, which are specialized text-to-text models, suffer from this limitation.

MonoT5 finetunes T5 on a binary classification task: given a query-document pair, the model is optimized to produce the words `\texttt{true}' if the document is relevant to the query and `\texttt{false}' otherwise. At inference time, the ranking score is obtained from the logits of the `\texttt{true}' and `\texttt{false}' tokens as follows:
\begin{align}
    s = \frac{e^{z_{\texttt{true}}}}{e^{z_{\texttt{true}}}+e^{z_{\texttt{false}}}}
\end{align}
where $z_{\tt true},z_{\tt false}$ are the logits of `\texttt{true}' and `\texttt{false}', respectively.

RankT5 directly learns to rank by optimizing a ranking-based loss function. This family of loss functions requires the model to directly output the ranking score for each query-document pair at training time, so that the unnormalized logit of a special unused token (`\texttt{extra\_id\_10}') in the vocabulary is used as ranking score. 

\begin{figure}[t]
    \centering
    \includegraphics[width=0.8\textwidth]{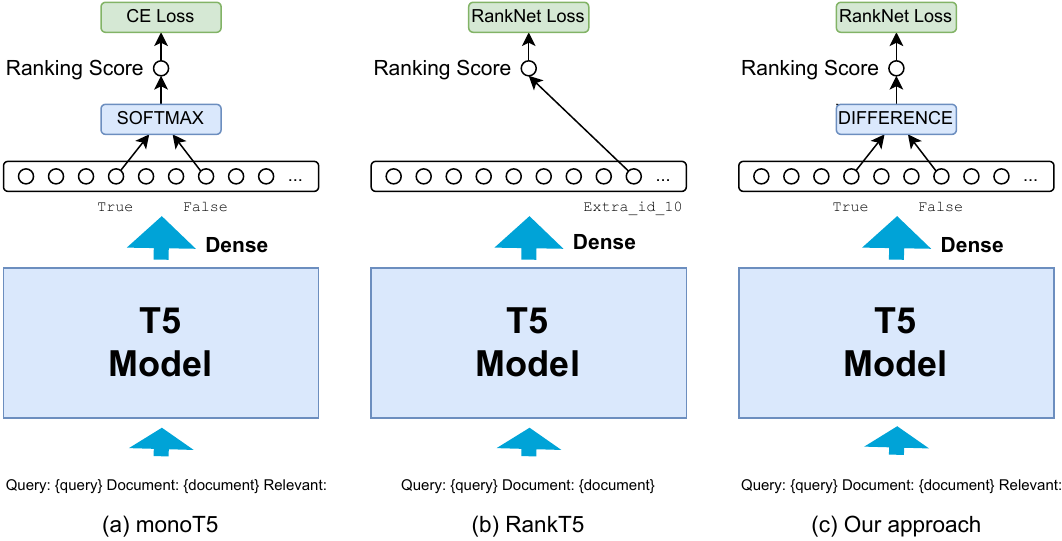}
    \caption{Illustration of the score strategies from monoT5, RankT5 and our proposed approach.}
    \label{fig:score-strategy}
\end{figure}

On one hand, monoT5 is not directly finetuned as a ranking model, which may not optimize its ranking performance. On the other hand, RankT5 does not exploit the learned representation in the language modeling head. To overcome both limitations, we propose a new approach. Our idea consists of using the difference between the unnormalized logits corresponding to the `\texttt{true}' and `\texttt{false}' tokens. In this way, the model is able to output a score directly at training time, and since it is optimized on top of the learned representations of the two tokens, we can make full use of the knowledge from the PLMs. An illustration of these scoring strategies is shown in Fig. \ref{fig:score-strategy}.

Accordingly, we adopt the RankNet loss~\cite{ranknet_loss}, a pairwise loss function that models the probability of one document being more relevant than another given a query. RankNet loss has shown compelling results in information retrieval~\cite{RankT5,rankgpt} and provided a solid foundation for our optimization process.

Given a query $q$ and $M$ passages ($p_1$, \dots, $p_M$) ordered by relevance $R=(r_1, \dots, r_M)$, where $r_i\in \{1, 2, \dots, M\}$ is the rank of the passage $p_i$ (if $r_i=1$, it means that $p_i$ is the most relevant passage), our model takes as input each query-document pair $(q, p_i)$ and outputs a relevance score $s_i$.

Therefore, we optimize our model with the following loss function measuring the correctness of relative passage orders:

\begin{align}
    \mathcal{L}_{RankNet} = \sum_{i=1}^M\sum_{j=1}^M \mathbb{I}_{r_i<r_j} \log(1+e^{s_i-s_j})
\end{align}

It should be noted that the adoption of different ranking loss functions could potentially lead to alternative outcomes, but exploring their potential differences is not the main purpose of this paper. Thus, we utilize the RankNet Loss here and leave the comparison of different ranking loss functions as future work.

\subsection{Distillation strategy}

Our distillation strategy aims to capture the reranking capability of LLMs, in our case ChatGPT, through constructing a query-document dataset. The core design principle is the synthesis of suitable artificial queries by query augmentation, and the subsequent use of multiple retrieval models and stages of distillation.

\paragraph{\bf Query Augmentation.}
Our query augmentation method is inspired by DRAG\-ON~\cite{DRAGON}, 
whereby two approaches are combined to amplify the size of training queries from a given corpus: sentence cropping and pseudo query generation. The former can readily scale up the size of query sets without any computationally expensive operation. The latter generates high-quality human-like queries using LLMs. The combination of the two strategies increases the diversity of the dataset and accordingly the challenge and complexity of the task.
%and has been utilized in~\cite{inpars,inparsv2} to adapt ranking models to specific datasets, where a reasonable amount of queries for training is often lacking.

Following DRAGON, we randomly sampled 10K queries from DRAGON's collection of cropped sentences, drawn from the MS MARCO corpus~\cite{MSMARCO}. Simultaneously, we sampled an additional 10K queries from the query pool created by docT5query~\cite{doc2query}, a specialized T5 model that generates queries based on a given passage.

\begin{figure}[t]
    \vspace{-0.5cm}
    \centering
    \includegraphics[width=0.85\textwidth]{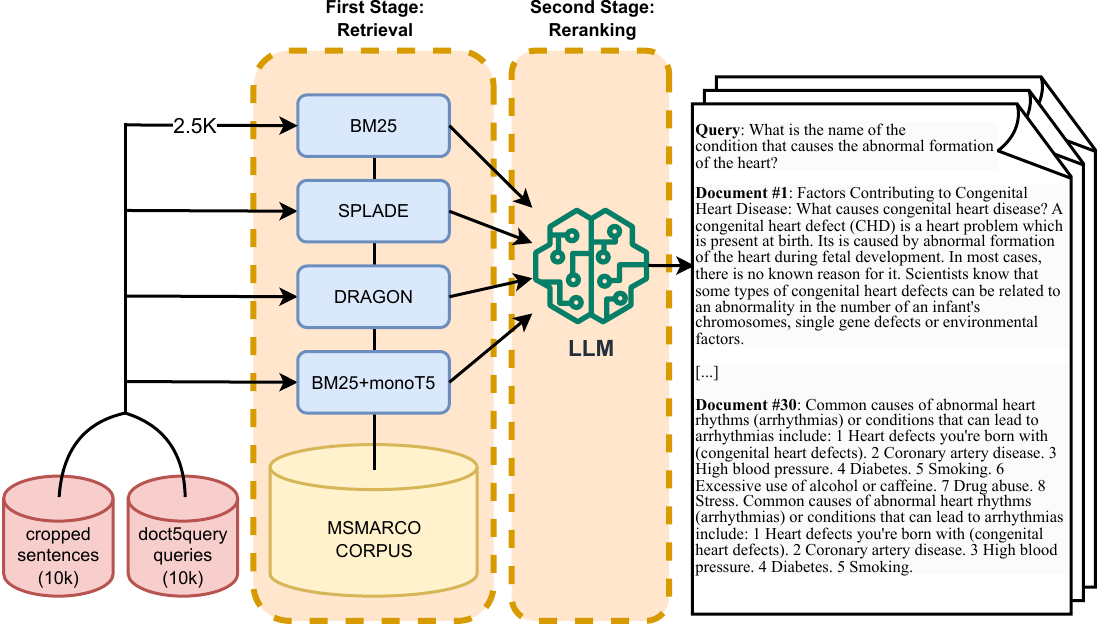}
    \caption{Illustration of the method used to build the dataset.}
    \label{fig:data-generation}
    %\vspace{-0.5cm}
\end{figure}

 %Our study found that a mixed query set, including cropped sentences and queries generated by a language model, proved to be the most effective approach. This aligns with the findings from the paper of DRAGON, thereby validating our choice of a balanced mixture of query types.

% The utilization of two different types of artificially generated queries introduces a layer of diversity to the tasks. This variety encourages a wider spectrum of responses from ChatGPT, furthering our understanding of its capabilities and potential weaknesses across a broader range of situations.

\paragraph{\bf First-stage distillation: retrieval.}
%This dataset is characterized by a two-tier distillation process, which incorporates an initial retrieval phase involving four distinct models, followed by a second phase where ChatGPT is employed. The selection of advanced retrieval models in the first phase is purposeful, %introducing a higher level of 
%enhancing the complexity of the dataset and providing a robust challenge for the subsequent use of ChatGPT.
The initial phase of our distillation process (see Fig. \ref{fig:data-generation}) involves splitting each of the two sets of 10K queries, one set composed of cropped sentences and another set composed of docT5query-generated queries, into four subsets of 2.5K queries each. To retrieve documents for these queries, we chose four distinct retrieval models:
\begin{itemize}
\item \textbf{BM25}~\cite{bm25}: A state-of-the-art bag-of-words approach that relies primarily on word overlap to match documents to queries. Consequently, its hard negatives are expected to challenge the language model on lexical-level matches.
\item \textbf{DRAGON}~\cite{DRAGON}: A dense retrieval model designed to detect semantic similarity between queries and passages. It pushes the language model towards understanding deeper semantic relations and contexts. We have chosen the DRAGON+ version.
\item \textbf{SPLADE}~\cite{splade,splade++,spladev2}: It serves as a kind of midpoint between BM25's focus on word overlap and DRAGON's emphasis on semantic similarity. It introduces a different level of complexity by considering interactions between the tokens of the document or query and all the tokens from the vocabulary. We have chosen the SPLADE++ version.
\item \textbf{monoT5}~\cite{monot5,monot5v2}: A combination of BM25 and monoT5 where the top-100 documents retrieved by BM25 are re-ranked using monoT5. It introduces negatives that are influenced by the ranking capabilities of a cross-encoder.
\end{itemize}

In all cases, we retrieve the top 30 documents for each query.

This methodology is designed to provide high-quality results and to diversify the types of challenges and contexts presented to ChatGPT in the subsequent distillation stage. 

Table \ref{table:intersection_sources} gives a quantitative account of the diversity of the documents retrieved by the four distinct models by computing the intersection rate between the sets of documents obtained from any two sources of supervision.\footnote{The average intersection rates were then calculated to provide a comprehensive view of the overall overlap among the retrieved documents from all sources:
\begin{align}
    \frac{1}{|\mathcal{Q}|}\sum_{q\in \mathcal{Q}} \frac{|S^1_q \cap S^2_q|}{N}
\end{align}
where $\mathcal{Q}$ is the whole query set, $S^1_q$ and $S^2_q$ represent the retrieved document set from two sources given the query $q$. This process was carried out separately for both types of queries: the cropped sentence queries and the docT5query-generated queries.}
The low mean intersection rates between different sources provide a clear evidence of the diversity among the retrieved document sets for both types of queries.

% \usepackage{graphicx}
\begin{table}[htb]
    \setlength{\abovecaptionskip}{-0.5cm}
\setlength{\belowcaptionskip}{0.25cm}
\caption{Average intersection rate between each pair of sources. The upper triangular part of the table represents the intersection rate for cropped sentences and the lower triangular part represents the intersection rate for docT5query-generated queries.}
\label{table:intersection_sources}
\centering
\scriptsize
% \resizebox{\textwidth}{!}{%
\begin{tabular}{l|cccc}
\hline \hline
doct5query \textbackslash \ cropped sentence (\%) & BM25             & SPLADE           & DRAGON           & monoT5           \\ \hline
BM25                               & \textbackslash{} & 20.0             & 29.0             & 49.8             \\
SPLADE                             & 17.8             & \textbackslash{} & 35.8             & 26.0             \\
DRAGON                             & 25.0             & 41.0             & \textbackslash{} & 38.4             \\
monoT5                             & 46.4             & 27.2             & 38.5             & \textbackslash{} \\ \hline \hline
\end{tabular}%
% }
\vspace{-.5cm}
\end{table}

\paragraph{ \bf Second-stage distillation: reranking.}
For reranking we used ChatGPT, in particular the checkpoint `\texttt{gpt-3.5-turbo-16k-0613}'. We prompted the model with the same prompt design proposed by RankGPT~\cite{rankgpt}, including all the 30 documents per query to rerank in a single prompt. 

We used the prompt made available by the RankGPT public repository. \footnote{\url{https://github.com/sunnweiwei/RankGPT}} We fed each of the 20K queries and their corresponding top 30 retrieved documents to ChatGPT, asking it to provide permutations of the indices of these documents, ordered according to their relevance to the associated query. 

This approach requires significant computational resources due to the complexity of the task and the vast number of queries and documents involved. However, the total cost of this reranking operation using the ChatGPT API amounted to \$212, demonstrating the feasible financial aspect of employing a large-scale language model in creating such a diverse and complex dataset. %\footnote{This data set will be publicly available after acceptance.}

\paragraph{ \bf Train-Validation split.} We divided the dataset into training and validation splits. We included 1,000 queries in the validation set: 500 queries generated by docT5query and the rest extracted as cropped sentences. The remaining 19,000 samples were allocated to the training set. 

\section{Experimental setup}
\label{sec:Experimental setup}

\subsection{Datasets}

We evaluate our approach using TREC-DL2019, TREC-DL2020 and BEIR for the zero-shot evaluation. 
All comparisons on TREC-DL2019 and TREC-DL2020 are based on the reranking of top-100 passages retrieved by BM25~\cite{bm25} for each query. %This is the same setting as existing work evaluating zero-shot LLM methods~\cite{rankgpt,prp}. 
The evaluation on the BEIR benchmark is based on the reranking of the top-100 passages retrieved by three different retrievers: BM25, SPLADE++~\cite{splade++}, and DRAGON+~\cite{DRAGON}. The objective is to evaluate the adaptability of the rerankers to different retrievers. We also present the evaluation by reranking the top-1000 documents retrieved by BM25, to give a broad view of the performances in different settings.

\subsection{Training}

We initialized our model with pretrained Flan-T5-xl checkpoint~\cite{flant5}. We set the maximum input sequence length to 500 as monoT5. The batch size is set to $32$, meaning that the parameters are updated 
%after computing the score for 
every $32\times 30$ query-document pairs. We utilize the AdamW~\cite{adamw} optimizer with a constant learning rate of $5e-5$. 

\subsection{Baselines}

We evaluate our model on the TREC-DL2019 and TREC-DL2020 competitions against several baselines including three supervised methods: monoBERT~\cite{monobert}, monoT5-3B~\cite{monot5,monot5v2}, RankT5~\cite{RankT5}; two zero-shot LLM-based methods: the \textit{listwise} prompting based approach of RankGPT~\cite{rankgpt}, using both \texttt{gpt-3.5-turbo} and \texttt{gpt-4}, and the sliding window approach performed only for 10 passes of PRP~\cite{prp}, using Flan-T5-xl (3B), Flan-T5-xxl (11B) and Flan-UL2 (20B); and a representative distilled model based on DeBERTav3 proposed in~\cite{rankgpt} as the only other LLM distillation method other than ours.
% \begin{itemize}
%     \item monoBERT~\cite{monobert};
%     \item monoT5-3B~\cite{monot5,monot5v2}
%     \item RankT5~\cite{RankT5}
% \end{itemize}
% We also consider the following zero-shot LLM-based baselines:
% \begin{itemize}
%     \item RankGPT~\cite{rankgpt}: the \textit{listwise} prompting based approach using both \texttt{gpt-3.5-turbo} and \texttt{gpt-4};
%     \item PRP~\cite{prp}: the sliding window approach performed only for 10 passes using Flan-T5-xl (3B), Flan-T5-xxl (11B) and Flan-UL2 (20B). 
% \end{itemize}
% We include in the comparison the distilled model based on DeBERTav3 proposed in~\cite{rankgpt} as the only other LLM distillaiton method other than ours.
Regarding the zero-shot evaluation on the BEIR benchmark, we evaluate our models comparing with three different rerankers, including InParsV2~\cite{inpars,inparsv2}, monoT5-3B~\cite{monot5,monot5v2}, and the distilled DeBERTav3 model proposed in~\cite{rankgpt}.
% \begin{itemize}
%     \item InParsV2~\cite{inpars,inparsv2};
%     \item monoT5-3B~\cite{monot5,monot5v2};
%     \item The distilled DeBERTav3 model proposed in~\cite{rankgpt}.
% \end{itemize}

\subsection{Results}
Our results on the TREC-DL2019 and TREC-DL2020 benchmarks are summarized in Table \ref{table:supervised_res}. Tables \ref{table:beir_result} and \ref{table:beir-1000} instead summarize respectively the results on the BEIR benchmark by reranking the top-100 and the top-1000 documents. 

% \usepackage{graphicx}
% \usepackage[normalem]{ulem}
\begin{table}[htb]
\setlength{\abovecaptionskip}{-0.cm}
\setlength{\belowcaptionskip}{0.5cm}
\caption{Results on TREC-DL2019 and TREC-DL2020 datasets by reranking top 100 documents retrieved by BM25. The column titled `\#Calls' indicates the exact number of inference times of LLM when reranking the top 100 documents. The `Input Size' column uses the notation $|q|+n|d|$: $|q|$ represents one query and $n|d|$ indicates the number of documents included. For instance, $|q|+20|d|$ signifies an input of one query with $20$ documents. Best model is highlighted in boldface and the second best is underlined for each metric. All the reported results apart from the LLM distillation Methods are taken from the original papers.}
\label{table:supervised_res}
\resizebox{\textwidth}{!}{%
\begin{tabular}{lllllcccccc}
\hline \hline
\multicolumn{1}{l|}{Method}         & \multicolumn{1}{l|}{LLM}           & \multicolumn{1}{l|}{Size}           & \multicolumn{1}{l|}{\#Calls} & \multicolumn{1}{l|}{Input Size}           & \multicolumn{3}{c|}{TREC-DL2019}                & \multicolumn{3}{c}{TREC-DL2020} \\ \hline
 
  \multicolumn{5}{l|}{} &
  \multicolumn{1}{c}{nDCG@1} &
  \multicolumn{1}{c}{nDCG@5} &
  \multicolumn{1}{c|}{nDCG@10} &
  \multicolumn{1}{c}{nDCG@1} &
  \multicolumn{1}{c}{nDCG@5} &
  \multicolumn{1}{c}{nDCG@10} \\ \hline
\multicolumn{1}{l|}{BM25}           & \multicolumn{1}{l|}{-} & \multicolumn{1}{l|}{-}& \multicolumn{1}{l|}{-}& \multicolumn{1}{l|}{-}            & 54.26       & 52.78 & \multicolumn{1}{c|}{50.58} & 57.72         & 50.67   & 47.96  \\ \hline
\multicolumn{11}{c}{Supervised Methods} \\ \hline
\multicolumn{1}{l|}{monoBERT}       & \multicolumn{1}{l|}{BERT}     & \multicolumn{1}{l|}{340M} & \multicolumn{1}{l|}{100} & \multicolumn{1}{l|}{$~|q|+|d|$}    & 79.07       & 73.25 & \multicolumn{1}{c|}{70.50} & 78.70         & 70.74   & 67.28  \\
\multicolumn{1}{l|}{monoT5}         & \multicolumn{1}{l|}{T5-xl}    & \multicolumn{1}{l|}{3B} & \multicolumn{1}{l|}{100} & \multicolumn{1}{l|}{$~|q|+|d|$}    & 79.07       & 73.74 & \multicolumn{1}{c|}{71.83} & 80.25         & 72.32   & 68.89  \\
\multicolumn{1}{l|}{RankT5}         & \multicolumn{1}{l|}{T5-xl}    & \multicolumn{1}{l|}{3B} & \multicolumn{1}{l|}{100}  & \multicolumn{1}{l|}{$~|q|+|d|$}   & 77.38       & 73.94 & \multicolumn{1}{c|}{71.22} & {\ul 80.86}         & 72.99   & 69.49  \\ \hline
\multicolumn{11}{c}{LLM distillation Methods}   \\ \hline
\multicolumn{1}{l|}{RankGPT}       & \multicolumn{1}{l|}{DeBertaV2}  & \multicolumn{1}{l|}{184M} & \multicolumn{1}{l|}{100} & \multicolumn{1}{l|}{$~|q|+|d|$} & 78.68       & 69.77 & \multicolumn{1}{c|}{66.56} &  59.26  &  59.83   & 59.43  \\
\multicolumn{1}{l|}{TWOLAR-large}       & \multicolumn{1}{l|}{Flan-T5-large}  & \multicolumn{1}{l|}{783M} & \multicolumn{1}{l|}{100} & \multicolumn{1}{l|}{$~|q|+|d|$} & 79.84       & 75.94 & \multicolumn{1}{c|}{72.82} & 79.94   & 71.35   & 67.61  \\ 
\multicolumn{1}{l|}{TWOLAR-xl}       & \multicolumn{1}{l|}{Flan-T5-xl}  & \multicolumn{1}{l|}{3B} & \multicolumn{1}{l|}{100} & \multicolumn{1}{l|}{$~|q|+|d|$} & 78.29       & {\ul 76.71} & \multicolumn{1}{c|}{{\ul 73.51}} & 80.25   & 73.73   & \bf 70.84  \\ \hline
\multicolumn{11}{c}{Zero-shot LLM Methods}  \\ \hline
\multicolumn{1}{l|}{RankGPT}        & \multicolumn{1}{l|}{gpt-3.5-turbo} & \multicolumn{1}{l|}{154B$^\ast$} & \multicolumn{1}{l|}{10} & \multicolumn{1}{l|}{$~|q|+20|d|$} & {\ul 82.17} & 71.15 & \multicolumn{1}{c|}{65.80} & 79.32         & 66.76   & 62.91  \\
\multicolumn{1}{l|}{RankGPT} &
  \multicolumn{1}{l|}{gpt-4} & \multicolumn{1}{l|}{1T$^\ast$} & \multicolumn{1}{l|}{2$^\dagger$} & \multicolumn{1}{l|}{$~|q|+20|d|$} &
  \textbf{82.56} &
  \textbf{79.16} &
  \multicolumn{1}{c|}{\textbf{75.59}} &
  78.40 &
  {\ul 74.11} &
  {\ul 70.56} \\
\multicolumn{1}{l|}{PRP-Sliding-10} & \multicolumn{1}{l|}{Flan-T5-xl} & \multicolumn{1}{l|}{3B} & \multicolumn{1}{l|}{990}  & \multicolumn{1}{l|}{$~|q|+2|d|$} & 75.58       & 71.23 & \multicolumn{1}{c|}{68.66} & 75.62         & 69.00   & 66.59  \\
\multicolumn{1}{l|}{PRP-Sliding-10} & \multicolumn{1}{l|}{Flan-T5-xxl} & \multicolumn{1}{l|}{11B} & \multicolumn{1}{l|}{990} & \multicolumn{1}{l|}{$~|q|+2|d|$} & 64.73       & 69.49 & \multicolumn{1}{c|}{67.00} & 75.00         & 70.76   & 67.35  \\
\multicolumn{1}{l|}{PRP-Sliding-10} &
  \multicolumn{1}{l|}{Flan-UL2} & \multicolumn{1}{l|}{20B} & \multicolumn{1}{l|}{990} & \multicolumn{1}{l|}{$~|q|+2|d|$}&
  78.29 &
  75.49 &
  \multicolumn{1}{c|}{72.65} &
  \textbf{85.80} &
  \textbf{75.35} & 
  70.46 \\ \hline \hline 
  \multicolumn{11}{l}{$^\ast$ OpenAI has not publicly released the amount of parameters and the numbers are based on public estimates \cite{gpt3.5vsgpt4} \cite{gpt4-review-opp}.} \\
  \multicolumn{11}{l}{$^\dagger$ In \cite{rankgpt}, gpt-4 reranks the top-30 passages reranked by gpt-3.5-turbo.}
 
\end{tabular}%
}
\vspace{-0.5cm}
\end{table}

% \usepackage{graphicx}
% \usepackage[normalem]{ulem}
\begin{table}[htb]
\setlength{\abovecaptionskip}{0.cm}
\setlength{\belowcaptionskip}{0.25cm}
\caption{Results on the BEIR Benchmark by reranking the top 100 documents with different retrievers. Best model is in boldface and second best is underlined for each dataset. Evaluation for InPars on CQADupStack is missing due to its unavailability on the Hugging Face hub. We computed statistical tests comparing our biggest model against the baselines. The results revealed no significant difference compared to InPars \( (p = 0.477) \), but indicated significant improvements over MonoT5-3B \( (p = 4.20 \times 10^{-4}) \) and RankGPT-Deberta \( (p = 2.82 \times 10^{-13}) \). It's noteworthy that while our model operates in a zero-shot manner, the InPars models have been fine-tuned for each BEIR dataset.}
\label{table:beir_result}
\resizebox{\textwidth}{!}{%
\begin{tabular}{l||c|ccccc||c|ccccc||c|ccccc}
\hline \hline
\multicolumn{1}{c||}{Retriever} &
  \multicolumn{6}{c||}{BM25} &
  \multicolumn{6}{c||}{SPLADE} &
  \multicolumn{6}{c}{DRAGON} \\ \hline
\multicolumn{1}{c||}{\rotatebox{280}{Reranker}} &
  - &
  \rotatebox{280}{MonoT5-3B} &
  \rotatebox{280}{InPars} &
  \rotatebox{280}{RankGPT-Deberta  } &
  \rotatebox{280}{TWOLAR-xl} &
  \multicolumn{1}{l||}{\rotatebox{280}{TWOLAR-large}} &
  - &
  \rotatebox{280}{MonoT5-3B} &
  \rotatebox{280}{InPars} &
  \rotatebox{280}{RankGPT-Deberta} &
  \rotatebox{280}{TWOLAR-xl} &
  \multicolumn{1}{l||}{\rotatebox{280}{{TWOLAR-large}}} &
  - &
  \rotatebox{280}{MonoT5-3B} &
  \rotatebox{280}{InPars} &
  \rotatebox{280}{RankGPT-Deberta} &
  \rotatebox{280}{TWOLAR-xl} & 
  \rotatebox{280}{TWOLAR-large} \\ \hline
  \multicolumn{19}{c}{nDCG@10} \\ \hline
TREC-COVID &
  59.5 &
  79.8 &
  82.5 &
  79.4 &
  82.7 & 
  84.3 & %%
  72.8 &
  82.9 &
  85.7 &
  80.1 & 
  85.2 &
  {\bf 86.9} & %%
  75.8 &
  82.8 &
  84.8 &
  82.6 &
  84.6 &
  {\ul 86.8} \\ %\cline{1-1}
NFCorpus &
  32.2 & 
  37.4 &
  35.0 &
  33.3 &
  36.6 &
  35.7 & %%
  34.8 &
  39.2 &
  38.8 &
  33.2 &
  37.3 &
  35.5 & %%
  33.9 &
  {\bf 39.7} &
  {\ul 39.3} &
  33.2 &
  37.9 &
  35.7 \\ %\cline{1-1}
FiQA-2018 &
  23.6 &
  46.1 &
  46.2 &
  32.7 &
  41.9 &
  41.1 & %%
  34.8 &
  50.0 &
  50.0 &
  33.7 &
  44.8 &
  43.8 & %%
  35.7 &
  {\bf 51.2} &
  {\ul 50.9} &
  43.1 &
  45.3 &
  44.8 \\ %\cline{1-1}
ArguAna &
  30.0 &
  33.4 &
  32.8 &
  21.1 &
  32.9 &
  34.7 & %%
  38.8 &
  31.7 &
  31.2 &
  18.6 &
  32.9 &
  34.6 & %%
  {\bf 46.9} &
  41.5 &
  40.9 &
  25.7 &
  42.8 &
  {\ul 45.5} \\ %\cline{1-1}
Tóuche-2020 &
  \textbf{44.2} &
  31.6 &
  29.6 &
  37.7 &
  37.1 &
  33.4 & %%
  24.6 &
  29.8 &
  28.7 &
  36.4 &
  35.2 &
  30.4 & %%
  26.3 &
  30.6 &
  29.4 &
  {\ul 38.2} &
  36.0 & 
  31.5 \\ %\cline{1-1}
Quora &
  78.9 &
  84.1 &
  84.8 &
  78.8 &
  87.2 &
  86.0 & %%
  83.5 &
  84.3 &
  85.1 &
  80.3 &
  {\ul 87.4} &
  86.0 & %%
  {\bf 87.5} &
  83.5 &
  84.4 &
  78.7 &
  87.2 & 
  85.7 \\ %\cline{1-1}
SCIDOCS &
  14.9 &
  19.0 &
  19.2 &
  16.1 &
  19.5 &
  18.3 & %% 
  15.9 &
  19.9 &
  {\bf 20.9} &
  16.4 &
  20.2 &
  18.8 & %%
  15.9 &
  19.8 &
  {\ul 20.7} &
  16.4 &
  20.2 & 
  18.8 \\ %\cline{1-1}
SciFact &
  67.9 &
  {\ul 76.4} &
  73.5 &
  70.5 &
  {\bf 76.5} &
  75.6 & %% 
  70.2 &
  {\ul 76.4} &
  76.0 &
  69.1 &
  75.6 &
  74.7 & %%
  67.8 &
  76.0 &
  75.7 &
  69.4 &
  75.6 &
  74.7 \\ %\cline{1-1}
NQ &
  30.6 &
  56.8 &
  57.8 &
  46.1 &
  58.0 &
  57.7 & %%
  53.7 &
  65.9 &
  66.4 &
  50.6 &
  {\ul 66.8} &
  65.8 & %%
  53.8 &
  65.1 &
  66.6 &
  50.6 &
  {\bf 66.9} &
  66.2 \\ %\cline{1-1}
HotpotQA &
  63.3 &
  74.2 &
  76.5 &
  69.9 &
  76.7 &
  75.9 & %%
  68.7 &
  74.1 &
  {\ul 77.1} &
  70.5 &
  {\bf 77.7} &
  76.4 & %%
  66.2 &
  72.9 &
  75.7 &
  69.8 &
  76.4 & 
  75.3 \\ %\cline{1-1}
DBPedia &
  31.8 &
  44.8 &
  44.0 &
  41.9 &
  48.0 &
  47.8 & %%
  43.6 &
  48.2 &
  51.1 &
  45.9 &
  {\bf 52.9} &
  51.6 & %%
  41.9 &
  47.2 &
  50.3 &
  44.9 &
  {\ul 52.1} &
  51.3 \\ %\cline{1-1}
FEVER &
  65.1 &
  83.2 &
  85.5 &
  80.2 &
  84.9 &
  83.4 & %%
  79.3 &
  85.0 &
  88.0 &
  81.8 &
  {\ul 87.5} &
  85.4 & %%
  78.0 &
  84.7 &
  {\bf 87.7} &
  81.7 &
  87.2 & 
  85.2 \\ %\cline{1-1}
Climate-FEVER &
  16.5 &
  27.4 &
  30.1 &
  24.2 &
  26.9 &
  26.1 & %%
  22.9 &
  28.7 &
  {\bf 32.8} &
  25.9 &
  28.9 &
  27.9 & %%
  22.7 &
  28.6 &
  {\ul 32.5} &
  25.9 &
  28.6 &
  27.4 \\ %\cline{1-1}
CQADupStack &
  30.2 &
  41.5 &
  - &
  34.7 &
  41.2 &
  40.6 & %%
  33.4 &
  43.7 &
  - &
  35.9 &
  43.6 &
  42.7 & %%
  35.4 &
  {\bf 44.4} &
  - &
  36.0 &
  {\ul 44.2} &
  43.4 \\ %\cline{1-1}
Robust04 &
  40.8 &
  56.6 &
  58.7 &
  52.8 &
  57.9 &
  58.3 & %%
  46.7 &
  62.1 &
  \textbf{64.3} &
  57.3 &
  64.9 &
  65.2 & %%
  48.1 &
  61.3 &
  {\ul 63.2} &
  56.6 &
  63.4 & 
  63.7 \\ %\cline{1-1}
Signal-1M &
  33.1 &
  32.2 &
  32.9 &
  {\ul 33.4} &
  \textbf{33.8} &
  33.9 & %% 
  30.0 &
  29.4 &
  30.3 &
  30.0 &
  30.1 &
  30.5 & %%
  30.0 &
  29.7 &
  30.4 &
  29.4 &
  30.2 &
  30.1 \\ %\cline{1-1}
BioASQ &
  52.3 &
  {\ul 57.2} &
  \textbf{59.8} &
  53.0 &
  56.2 &
  56.0 & %%
  49.7 &
  54.1 &
  57.2 &
  49.5 &
  54.6 &
  53.8 & %%
  43.4 &
  51.9 &
  54.4 &
  48.0 &
  51.9 &
  50.8 \\ %\cline{1-1}
TREC-NEWS &
  39.5 &
  48.5 &
  49.8 &
  51.8 &
  52.7 &
  50.8 & %%
  41.5 &
  50.0 &
  50.9 &
  {\ul 53.4} &
  53.3 &
  50.7 & %%
  44.4 &
  49.5 &
  50.8 &
  52.1 &
  {\bf 53.8} &
  50.0 \\ \hline
  \multicolumn{19}{c}{avg nDCG@10} \\ \hline
\multicolumn{1}{l||}{BEIR 18} &
  41.9 &
  51.7 &
  - &
  47.6 &
  52.8 &
  52.2 & %%
  46.9 &
  53.0 &
  - &
  48.3 &
  {\ul 54.4} &
  53.4 & %%
  47.4 &
  53.4 &
  - &
  48.5 &
  {\bf 54.7} &
  53.7 \\ 
\multicolumn{1}{l||}{BEIR 17} &
  42.6 &
  52.3 &
  52.9 &
  48.4 &
  53.5 &
  52.9 & %%
  47.8 &
  53.5 &
  55.0 &
  49.0 &
  55.0 &
  54.0 & %%
  48.1 &
  53.9 &
  {\ul 55.2} &
  49.2 &
  {\bf 55.3} &
  54.3 \\ \hline \hline

\end{tabular}%
}
\vspace{-0.5cm}
\end{table}

% \usepackage{graphicx}
% \usepackage[normalem]{ulem}
\begin{table}[hbt]
\centering
\caption{Results on the BEIR Benchmark by reranking the top 1000 BM25 retrieved documents. The best model is highlighted in boldface, and the second best is underlined for each dataset. All results, apart from TWOLAR-xl, are from~\cite{inparsv2}.}
\scriptsize
\label{table:beir-1000}
%\resizebox{\textwidth}{!}{%
\begin{tabular}{l||c|cccc}
\hline \hline
 & BM25 & monoT5-3B & InPars-v2 & RankT5 & TWOLAR-xl \\ \hline
 \multicolumn{6}{c}{nDCG@10} \\ \hline
TREC-COVID &
  59.5 &
  80.1 &
  {\bf 84.6} &
  82.3 &
  {\ul 84.3} \\ %\cline{1-1}
NFCorpus &
  32.2 &
  38.3 &
  {\ul 38.5} &
  {\bf 39.9} &
  37.3 \\ %\cline{1-1}
FiQA-2018 &
  23.6 &
  {\bf 50.9} &
  {\ul 50.9} &
  49.3 &
  45.2 \\ %\cline{1-1}
ArguAna &
  30.0 &
  {\ul 37.9} &
  36.9 &
  {\bf 40.6} &
  32.7 \\ %\cline{1-1}
Tóuche-2020 &
  {\ul 44.2} &
  30.9 &
  29.1 &
  {\bf 48.6} &
  35.9 \\ %\cline{1-1}
Quora &
  78.9 &
  83.5 &
  {\ul 84.5} &
  81.9 &
  {\bf 87.3} \\ %\cline{1-1}
SCIDOCS &
  14.9 &
  19.7 &
  {\bf 20.8} &
  19.1 &
  {\ul 20.3} \\ %\cline{1-1}
SciFact &
  67.9 &
  {\bf 77.4} &
  {\ul 77.4} &
  76.0 &
  76.8 \\ %\cline{1-1}
NQ &
  30.6 &
  62.5 &
  63.8 &
  {\bf 64.7} &
  {\ul 64.2} \\ %\cline{1-1}
HotpotQA &
  63.3 &
  76.0 &
  {\ul 79.1} &
  75.3 &
  {\bf 79.5} \\ %\cline{1-1}
DBPedia &
  31.8 &
  47.2 &
  {\ul 49.8} &
  45.9 &
  {\bf 52.0} \\ %\cline{1-1}
FEVER &
  65.1 &
  84.8 &
  {\bf 87.2} &
  84.8 &
  {\ul 86.7} \\ %\cline{1-1}
Climate-FEVER &
  16.5 &
  {\ul 28.8} &
  {\bf 32.3} &
  27.5 &
  27.8 \\ %\cline{1-1}
CQADupStack &
  30.2 &
  {\bf 44.9} &
  {\ul 44.8} &
  - &
  43.8 \\ %\cline{1-1}
Robust04 &
  40.8 &
  61.5 &
  {\ul 63.2} &
  - &
  {\bf 64.2} \\ %\cline{1-1}
Signal-1M &
  {\bf 33.1} &
  30.2 &
  30.8 &
  {\ul 31.9} &
  31.5 \\ %\cline{1-1}
BioASQ &
  52.3 &
  56.6 &
  {\bf 59.5} &
  {\ul 57.9} &
  56.0 \\ %\cline{1-1}
TREC-NEWS &
  39.5 &
  47.7 &
  {\ul 49.0} &
  - &
  {\bf 53.2} \\ \hline
  \multicolumn{6}{c}{avg nDCG@10} \\ \hline
 BEIR 18 &
  41.9 &
  53.3 &
  {\bf 54.5} &
  - &
  {\ul 54.4} \\ 
 BEIR 15 &
  42.9 &
  53.7 &
  {\ul 54.9} &
  {\bf 55.0} &
  54.5 \\ \hline \hline

\end{tabular}%
% }

\vspace{-0.5cm}
\end{table}

\section{Discussion}
\label{sec:Discussion}

In this section, We will delve into a comprehensive discussion of our experimental results, and in the following section, we will explore the ablation study in detail.

\paragraph{\bf On TREC-DL.}
In our evaluation on TREC-DL2019 and TREC-DL2020, our model demonstrated outstanding performance, consistently outperforming established supervised methods and LLM-distilled baselines. When set against zero-shot LLM baselines, our model either matches or exceeds their performance. Although the results are not directly comparable because we are using a specific checkpoint, we find it remarkable that our model outperforms even the teacher LLM used for the distillation process, i.e. \texttt{gpt-3.5-turbo}. We take it as an indication that our distillation strategy is well conceived. The sole model that distinctly outperformed ours was \texttt{gpt-4}. This performance difference suggests that leveraging a more advanced LLM for distillation within our methodology might lead to even superior outcomes. Importantly, this is achieved with significantly reduced computational overhead during inference since we distilled LLMs to obtain a much smaller task-specific model. It is worthwhile noticing the difference in size between the models used for comparison with TWOLAR. Remarkably the largest of the TWOLAR models is several orders of magnitude smaller than the largest RankGPT model. 

\paragraph{\bf On BEIR Benchmark.}
In our evaluations using the BEIR benchmark, TWO\-LAR consistently outperformed most existing baselines when reranking the top-100 documents, as shown in Table \ref{table:beir_result}. This is particularly significant when compared with the approach of models such as InPars.  InPars employs a strategy of fine-tuning a monot5-3B on generated, topic-specific data tailored for each of the 18 datasets within the BEIR benchmark. This strategy means that, for each dataset, their model has been exposed to data related to the topic in question. In contrast, TWOLAR has never been exposed to any topic-specific data, making it genuinely zero-shot when facing new topics and tasks. Furthermore, our method does not require fine-tuning for different applications and is thus more economical. 

It is worthwhile noticing the performance variations across different datasets within BEIR. In datasets with a specific focus, such as BioASQ, InPars tends to perform better due to its targeted fine-tuning on artificial topic-specific data. However, in datasets where queries are centered around general knowledge, like DBpedia entity, TWOLAR demonstrates a clear advantage over InPars. This highlights the robustness of TWOLAR's topic-agnostic approach and its applicability in a broad range of scenarios.

When reranking the top-1000 documents (Table \ref{table:beir-1000}), our model did not perform as well as when reranking the top-100 documents. However, the difference with the best performing model on BEIR 18 is minor (54.4 vs 54.5) and TWOLAR-xl outperforms every competitor in 5 out of 18 tasks. A possible explanation for this result lies in our model's training setup. Since our method optimizes for reranking a subset of 30 documents, it seems plausible that it can easily scale up to 100 documents, less easily to 1000. 

% \usepackage{graphicx}
\begin{table}[htb]
\setlength{\abovecaptionskip}{0.cm}
\setlength{\belowcaptionskip}{0.5cm}
\centering
\scriptsize
\caption{We perform ablation studies on the eight smallest datasets in BEIR benchmark. The reported scores are nDCG@10. In comparing our scoring strategy against the RankT5 scoring strategy, the statistical tests yielded a p-value of \( p = \text{0.268} \). *COV: TREC-COVID, SCI: SciFact, NFC: NFCorpus, TOU: Tóuche-2020, DBP: DBPedia, ROB: Robust04, SIG: Signal-1M, NEW: TREC-NEWS. }
\label{table:ablation}
% \resizebox{\textwidth}{!}{%
% \begin{tabular}{l|cccccccc|c}
\begin{tabular}{l|P{1cm}P{1cm}P{1cm}P{1cm}P{1cm}P{1cm}P{1cm}P{1cm}|P{1cm}}
\hline \hline
% \multicolumn{1}{l|}{} & TREC-COVID & SciFact & NFCorpus & Tóuche-2020 & DBPedia & Robust04 & Signal-1M & \multicolumn{1}{l|}{TREC-NEWS} & avg nDCG@10 \\ \hline \hline

 & COV & SCI & NFC & TOU & DBP & ROB & SIG & NEW & avg \\ 
 \hline \hline

Score Strategy & \multicolumn{9}{c}{effectiveness of score strategy - 19K train samples}                                            \\ \hline
Difference & 74.0     & 67.9     & \textbf{31.9} & \textbf{35.7} & \textbf{38.8} & \textbf{47.4} & 32.5     & \textbf{43.7} & \textbf{46.5} \\
RankT5   & \textbf{74.1} & \textbf{69.2} & 31.5     & 32.2     & 36.2     & 47.0     & \textbf{34.1} & 41.2     &     45.7 \\ \hline \hline
\# documents & \multicolumn{9}{c}{effectiveness of amount of documents - 19K train samples}                                       \\ \hline
30         & \textbf{74.0} &     67.9 &     31.9 & \textbf{35.7} & \textbf{38.8} &     47.4 & \textbf{32.5} & \textbf{43.7} & \textbf{46.5} \\
20         &     73.0 & \textbf{69.8} & \textbf{32.5} &     31.7 &     37.9 & \textbf{47.5} &     31.9 &     40.8 &     46.3 \\
10         &     72.3 &     65.6 &     29.6 &     28.4 &     34.2 &     43.2 &     30.4 &     37.9 &     42.7 \\ \hline \hline
Not used source & \multicolumn{9}{c}{effectiveness of first source of supervision - $\sim$14K train samples}                         \\ \hline
- BM25     &     72.7 &     70.3 &     31.8 &     31.8 &     37.7 &     47.0 &     31.9 &     41.5 &     45.6 \\
- SPLADE   &     73.9 & \textbf{70.9} & \textbf{33.6} &     32.7 & \textbf{38.6} & \textbf{48.8} &     32.4 &     42.5 & \textbf{46.3} \\
- DRAGON   & \textbf{74.0} &     67.7 &     32.9 & \textbf{33.9} &     37.6 &     47.7 & \textbf{33.1} &     43.2 &     46.2 \\
- monoT5   &     73.9 &     69.4 &     31.8 &     33.0 &     36.3 &     46.6 &     32.5 & \textbf{43.6} &     45.9 \\ \hline \hline
Type of query & \multicolumn{9}{c}{effectiveness of type of query - 9.5K train samples}                                            \\ \hline
Mixed      & \textbf{75.5} &     67.3 &     30.4 & \textbf{34.0} &     37.2 & \textbf{46.2} & \textbf{31.8} &     41.6 & \textbf{45.5} \\
Sentence   &     67.2 & \textbf{67.9} & \textbf{31.4} &     32.7 &     32.2 &     44.8 &     31.7 &     39.7 &     43.4 \\
docT5query &     74.6 &     59.4 &     31.2 &     33.4 & \textbf{37.8} &     44.9 &     28.1 & \textbf{44.0} &     44.2 \\ \hline \hline
\end{tabular}%
% }
\vspace{-0.5cm}
\end{table}
\subsection{Ablation study}

We conducted an extensive ablation study in order to validate our design choices. Due to computational constraints, these experiments were performed using the smaller \texttt{flan-t5-small} checkpoint, with 77M parameters. Furthermore, we evaluated the models on reranking the top-100 documents retrieved by BM25 from a subset of 8 smallest datasets from the BEIR benchmark, including TREC-COVID, SciFact, NFCorpus, Tóuche-2020, DBPedia, Robust04, Signal-1M, and TREC-NEWS. 

The results are summarized in Table \ref{table:ablation}.

\paragraph{ \bf Scoring Strategy Effectiveness.}
Our scoring strategy achieved an average nDCG@10 of 46.5, showing superior performance compared to 45.7 for the RankT5 scoring approach, which indicates the importance of properly exploiting the knowledge from the language modeling head of PLMs. We did not make a direct comparison with the softmax method used in monoT5, due to the inherent differences in the pipeline structure: while our method and the RankT5 method allow for direct finetuning of the model to rank documents, the monoT5 approach operates on a fundamentally different mechanism, making a direct comparative analysis less feasible and potentially misleading in evaluating the distinct methodologies.
% We compared our proposed scoring strategy, which utilizes the difference between the '\texttt{true}' and '\texttt{false}' logits, with the strategy used in RankT5, based on the logit of an extra token in the T5 vocabulary. Our analysis indicated superior performance for our proposed strategy, achieving an average nDCG@10 of 46.5, in comparison to 45.7 for the RankT5 scoring approach.

\paragraph{ \bf Documents per training samples.}
We trained models with varying numbers of documents per training sample. Our results suggest a clear advantage in using more than 10 documents per sample. The trade-off between 20 and 30 is less clear, with nDCG@10 scores of 46.3 and 46.5 respectively, suggesting diminishing returns beyond 20 documents for the top-100 reranking task.

\paragraph{ \bf Effectiveness of first source of supervision.}
We conducted four individual experiments by excluding each source of supervision (BM25, SPLADE, DRAGON, monoT5) from the training set and training the model on the residual data. This allowed us to evaluate the individual contribution of each retrieval strategy to the overall performance of the model.

Our results demonstrate that BM25, even being a traditional bag-of-words method, still plays a critical role in the model's performance. This result may be also due to the fact that, following standard practice, during the test the top-100 documents have been retrieved using BM25 itself. Conversely, when SPLADE and DRAGON were excluded during training, the performance drop was not substantial, which suggests that the main contribution comes from blending lexical and semantic models rather than including multiple and possibly equivalent semantic models. 

\paragraph{ \bf Impact of Query Type.}
We also trained models exclusively on cropped sentences, docT5query generated queries, and a mixed subset of both types. The model trained only with docT5query generated queries, which are formulated as natural language questions, had a higher average performance than the model trained only on cropped sentences. This suggests that training with grammatically correct questions is more important.

Interestingly, for datasets where the queries were predominantly formed as `what' or `how' questions, such as TREC-COVID, the model trained on docT5query queries delivered a strongly superior performance. Conversely, the model trained with cropped sentences performed better in specific datasets where the queries are not expressed as a question.
%in natural language
For example, the queries in SciFact are expert-written claims, aiming to find evidence in annotated abstracts. Here, the model trained with cropped sentences achieved an nDCG@10 score of 67.9, significantly outperforming the model trained with docT5query queries, which scored 59.4. 

When we trained the model on a mixed subset comprising an equal proportion of both query types, it exhibited the best overall performance. This highlights the benefit of a diverse training regimen incorporating natural questions (docT5query) and sentences cropped directly from documents.

In summary, these results of the ablation study underscore the value of our proposed scoring strategy, the importance of incorporating sufficient documents per training sample, the significant contribution of BM25 as a supervision source, and the advantages of a mixed query approach.

% \vspace{-1.cm}

\section{Conclusion}
\label{sec:Conclusion}

The paradigm shift, enabled by LLMs, suggests that traditional methods relying heavily on handcrafted labeled data might no longer be the most effective or efficient approach for certain machine learning tasks. Indeed, as LLMs continue to showcase their prowess, there is a promising realization that they can be harnessed to provide the needed supervision, reducing the need for manual data labeling. However, tasks that demand efficiency, such as information retrieval, often cannot deploy LLMs directly due to their substantial computational overhead. In such scenarios, distillation enables the retention of the LLM's capabilities in a more computationally amenable format.

In this paper, we presented a novel two-step LLM-augmented distillation approach for passage reranking. Our method capitalizes on the strengths of LLMs to enable computationally efficient information retrieval systems, with performance comparable or even superior to that of state-of-the-art baselines and a reduction in size by several orders of magnitude. Our experiments, conducted across various benchmarks, demonstrate robustness and generality of our approach across domains. An ablation offers further insight about the crucial elements of our architectural design.

%This dual advantage of superior performance and efficiency positions our method as a compelling benchmark for future reranking tasks.

%underlining its robustness and effectiveness.

Looking forward, TWOLAR offers promising avenues for scalability. In the future, we plan to further our experimentation by substituting the 3B model with an 11B version, expanding the number of queries, increasing the sources of supervision, or even refining the quality of the LLM used for distillation, for example by experimenting with more powerful generative language models.

\bibliography{biblio}
\bibliographystyle{splncs04}

%

% \printbibliography[heading=bibintoc] % biblatex

\end{document}